\documentclass[twoside]{ilcws08}
\usepackage[latin1]{inputenc}
\usepackage[dvips]{graphicx,epsfig,color}
\usepackage{wrapfig,rotating}
\usepackage{amssymb,amsmath,array}

\begin{document}
\title{
Summary of One Year Operation of the EUDET CMOS Pixel Telescope} 
\author{Ingrid-Maria Gregor$^1$
\thanks{This work is supported by the Commission of the European Communities
under the 6$^{th}$ Framework Programme ¨Structering the European Research 
Area¨, contract number RII3-026126.}
\vspace{.3cm}\\
1- DESY Hamburg, Notkestr. 85, D-22607 Hamburg, Germany
}

\maketitle

\begin{abstract}
Within the EUDET consortium a high resolution pixel beam telescope is
being developed. The telescope consists of up to six planes of monolithic
active pixel sensors.
A flexible data acquisition environment is available for the telescope and
the system is equipped with all the required infrastructure. Since the
first installation of a demonstrator telescope in 2007, it has been
extensively tested and used by various detector R\&D groups. The results 
of test beam measurements are described here, demonstrating the telescope 
performance. 

\end{abstract}

\section{Introduction}
\begin{wrapfigure}{r}{0.55\columnwidth}
\centerline{\includegraphics[width=0.5\columnwidth]{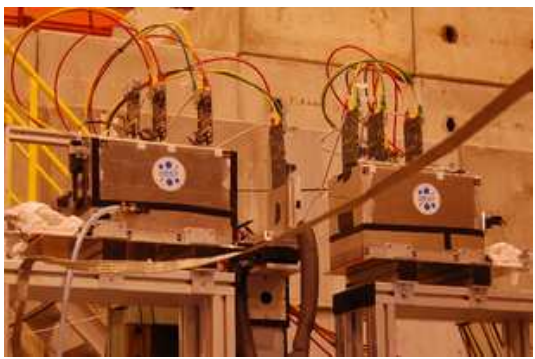}}
\caption{The EUDET pixel telescope installed at the CERN hadron
test beam.}\label{Fig:MV}
\end{wrapfigure}

\label{sec:introduction}
The EUDET project~\cite{ref:eudet}, 
which is supported by the EU in the 6th Framework Programme (FP6), aims 
to provide infrastructure for the R\&D of detector technologies towards 
the international linear collider. Within the EUDET project the JRA1 activity 
works on the improvement of test beam infrastructure. For this purpose, a high 
resolution pixel telescope is being developed. The design goals include a 
high position resolution ($\sigma < 3.0~\mu$m) and readout rate of 
1~kHz. 

The construction of the telescope is performed in two steps. In June 
2007, the so-called demonstrator telescope was installed for the first
time using an analog readout. After the first successful
operation at the electron beam at DESY, the demonstrator was
transported to CERN and its performance was studied using 180~GeV
hadrons at the SPS~\cite{ref:toto_hawaii_2007}. After the first
successful integration of a Device Under Test (DUT) in September 
2007~\cite{ref:depfet_first_dut}, the demonstrator telescope has been used
by various groups and was improved continuously. 

\section{Description of the Telescope} 

\subsection{The sensor}
\label{sec:sensor}
The MimoTEL sensor, used for the demonstrator telescope, is a Monolithic
Active Pixel Sensor (MAPS) produced in the AMS 0.35 OPTO
process. It is subdivided in four
sub-arrays of 64$\times$256 pixels with a pixel
pitch of 30$\times$30~$\mu$m$^{2}$, resulting in a sensor size of
7.7$\times$7.7~cm$^{2}$. A high resolution sensor with a pitch of
10~$\mu$m can be located close to the DUT to further increase the 
resolution. 

\subsection{The DAQ system}
\label{sec:daq}
The DAQ system can be summarised as follows: All data from the sensors
is transferred via frontend boards to an intermediate readout and data
reduction board called EUDRB (EUDET Data Reduction Board)~\cite{ref:eudrb}. 
The EUDRB board allows the first steps of the data 
processing online to be performed. Two I/O busses are supported: For the
telescope the VME64x bus is used to allow high speed
data transfer and synchronous operation with other devices while an
USB2.0 interface is foreseen for standalone testing. A mother / daughter
board scheme has been followed to maximise the flexibility. All
computing and memory elements are located on the motherboard while the
sensor specific components have been implemented on removable and
interchangeable daughter cards. 

Another important component of the DAQ system is the trigger logic unit
(TLU)~\cite{ref:tlu}. It is considered as the replacement for a NIM crate and
can generate any coincidence or anticoincidence of four trigger
scintillators. Six LVDS and two TTL interfaces are
provided. Furthermore, the TLU generates event numbers and time
stamps. It is connected by USB2.0 to a control PC running the Linux
operating system that is in turn connected to the main DAQ PC through
gigabit ethernet.

A custom DAQ system named EUDAQ has been implemented in C++~\cite{ref:eudaq}. 
Several producer tasks communicate with a global run control using sockets. 
These producer tasks connect to the hardware of the beam telescope, to the TLU and
eventually to the DUT. Data from all producers is sent to the central
data collector and can be monitored by several processes. An online
monitor, based on the ROOT framework, shows online data quality
monitoring histograms and a process to collect log messages is 
available. EUDAQ runs on MacOS, Linux and Windows using cygwin.

\subsection{The offline analysis software}
\label{sec:software}

For the offline reconstruction of track positions in the DUT the
software package EUTelescope~\cite{ref:eutelescope} has been developed, which 
is implemented as a set of Marlin processors~\cite{ref:ilcsoft}. This design 
allows to integrate the DUT data at different steps of the analysis chain. 
Furthermore, the package can be executed on the Grid to allow a fast 
processing of  large datasets.

\section{Test beam results}
\label{sec:results}
\begin{figure}[ht]
\begin{center}
\includegraphics[width=11cm]{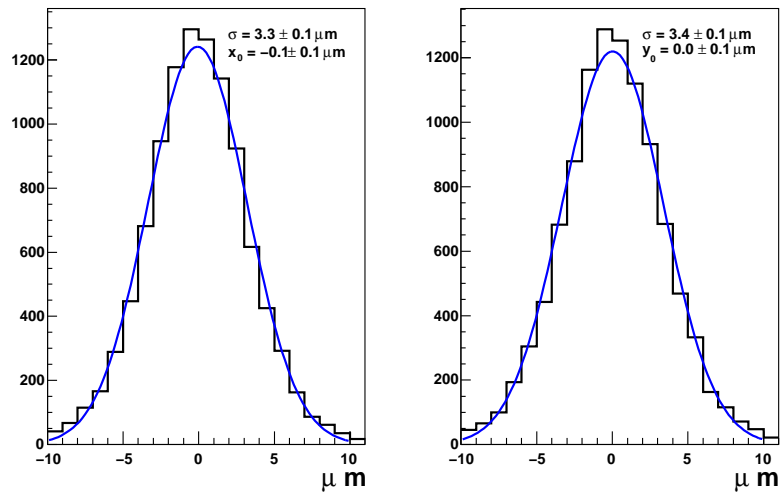}
\caption{\label{fig:line_fit}Residuals in the middle telescope
  sensor in the $X$ (left) and $Y$ (right) directions. This sensor was
  excluded from the track fit and hence acts as DUT. The data was recorded
at the CERN SPS hadron test beam.}
\end{center}
\end{figure}

Figure~\ref{fig:line_fit} shows
the residuals of the tracks in the middle out of 5 sensor
planes. Here the middle telescope sensor acts as DUT while the other
planes are used to predict the track positions in the DUT. The
observed width is consistent with the expectation for the given
telescope geometry assuming a position resolution of
3.0~$\mu$m for the DUT as well as for the other sensors used to fit
tracks. Also, measurements at DESY using 3 and 6~GeV electrons based on
an extrapolation to infinite energy are in agreement with this sensor 
resolution. The telescope was used by 7 different detector R\&D groups 
for resolution studies of their own systems. All of them were satisfied
with the telescope performance and are planning to use it again in 2009.

\section{Summary}
\label{sec:summary}

The demonstrator telescope has been running successfully during several
test beam measurements for about one year. Modularity was one of the
most important design aspects for the DAQ hardware and software as
well as for the offline analysis package. The analysis of test beam
data shows that the performance of the demonstrator fulfils the
expectations. An increased active area and zero suppression on the
sensors will be offered by the final telescope which is under construction. 
Groups interested in using the device are welcome to contact the 
EUDET consortium.



\begin{footnotesize}


\end{footnotesize}


\end{document}